\documentclass[12pt]{JHEP3}
\usepackage{amsmath,epsfig}
\def\be{\begin{equation}}
\def\ee{\end{equation}}
\def\ba{\begin{eqnarray}}
\def\ea{\end{eqnarray}}

\def\be{\begin{equation}}
\def\ee{\end{equation}}
\def\bea{\begin{eqnarray}}
\def\eea{\end{eqnarray}}

\def\yzero{\smash{\hbox{$y\kern-4pt\raise1pt\hbox{${}^\circ$}$}}}

\def\beq{\begin{equation}}
\def\eeq{\end{equation}}
\def\beqa{\begin{eqnarray}}
\def\eeqa{\end{eqnarray}}

\def\-{\hphantom{-}}

\def\s2{\frac{1}{\sqrt2}}

\def\beq{\begin{equation}}
\def\eeq{\end{equation}}
\def\beqa{\begin{eqnarray}}
\def\eeqa{\end{eqnarray}}

\def\IF{\relax{\rm I\kern-.18em F}}
\def\II{\relax{\rm I\kern-.18em I}}
\def\IP{\relax{\rm I\kern-.18em P}}
\def\IC{\relax\hbox{\kern.25em$\inbar\kern-.3em{\rm C}$}}
\def\IR{\relax{\rm I\kern-.18em R}}

\def\Dsl{\,\raise.15ex\hbox{/}\mkern-13.5mu D} 
\def\IZ{Z\kern-.4em  Z}

\title{SL(2,Z) symmetries, Supermembranes and Symplectic Torus Bundles}

\author{M. P. Garc\'\i a del Moral $^1$,I. Mart\'\i n, $^2$ J. M. Pe\~na $^3$, A. Restuccia $^4$\,\footnote{E-mail:
\emph{garciamormaria@uniovi.es;isbeliam@usb.ve;jmpena@ciens.fisica.ucv.ve;arestu@usb.ve}}\\
$^1$ Departamento de F\'\i sica, Universidad de Oviedo, Avda Calvo
Sotelo S/n. Oviedo, Espa\~na\\
$^2$ Departamento de F\'\i sica, Universidad Sim\'on Bol\'\i var\\
Apartado 89000, Caracas 1080-A, Venezuela
\\
$^3$ Departamento de F\'\i sica, Facultad de Ciencias,\\
 Universidad Central de Venezuela,
 A.P. 47270, Caracas 1041-A, Venezuela\\ 
$^4$ Departamento de F\'\i sica, Universidad de Antofagasta, Aptdo 02800, Chile \\
 $\&$ Departamento de F\'\i sica, Universidad Sim\'on Bol\'\i var\\
Apartado 89000, Caracas 1080-A, Venezuela}

\abstract{ We give the explicit formulation of the 11D supermembrane as a
symplectic torus bundle with non trivial monodromy and non vanishing
Euler class. This construction allows a classification of all supermembranes showing explicitly the discrete $ SL(2,Z)$ symmetries associated to dualities. 
It hints  as the origin in M-theory of the gauging of the effective theories associated to string theories. }

\preprint{FPAUO-11/06}


\begin{document}
\section{Introduction}
Nonperturbative effects like monopoles, instantons in conventional gauge theories, or dualities in the context
of M/string theories rely on global aspects  of those theories.
Properties like confinement  may well be due to
non trivial topological aspects as well. Non trivial fibrations have also been used in the context of noncommutative
theories, like the noncommutative formulation of the torus \cite{ho}
as well to characterize compactification spaces useful for string
phenomenology, see for example \cite{ovrut},\cite{ralph}. 

There is evidence that string theory can be consistently defined in
non-geometric backgrounds in which the transition functions between
coordinate patches involve not only diffeomorphisms and gauge
transformations but also duality transformations \cite{7samtleben}\cite{3hull99}.
Such backgrounds can arise from compactifications with duality
twists \cite{11dabholkarhull} or from acting on geometric
backgrounds with fluxes with T-duality \cite{3hull99},
\cite{4hull04}, \cite{5wetch} or mirror symmetry. In special cases,
the compactifications with duality twists are equivalent to
asymmetric orbifolds which can give consistent string backgrounds
\cite{12flournoyw}, \cite{13kawai}, \cite{hulldabholkar}. In this
type of compactifications, T-folds are constructed by using  strings formulated on a doubled torus $T^{2n}$ with
n coordinates  conjugate to the momenta and the other n coordinates 
conjugate to the winding modes \cite{4hull04}.

In \cite{3hull99,4hull04} it was argued that a fundamental
formulation of string/M-theory should exist in which the T-
and U-duality symmetries are manifest from the start. The natural framework for M-Theory would generalize ten- or
eleven-dimensional space-time into a higher-dimensional geometry in
which auxiliary dimensions would be related to non-metric degrees of
freedom. The duality symmetries of string- and M-theory would be
discrete geometric symmetries of this generalized space. In
particular, it was argued that many massive, gauged supergravities
cannot be naturally embedded in string theory without such a
framework \cite{5wetch}, \cite{6hullreid}, \cite{7samtleben}.
Examples of generalized T-folds can be obtained by constructing torus fibrations over base manifolds with
non-contractible cycles. In
particular, an example  is to consider $S^1$ as base manifold when the monodromy of the theory in the fibres
includes a non-geometric element of O(Z) corresponding to a
generalised T-duality \cite{reids}. However, in spite of advances, up to our knowledge, a full-fledged realization of these ideas
in terms of the string worldsheet action for String Theory , or
in terms of the supermembrane for M-theory is still lacking.

The aim of this paper is to prove that the action of the Supermembrane  with nontrivial central charges, whose local structure was given in \cite{torrealba}\cite{ovalle}\cite{sl2z} 
may be globally defined in terms of sections of a symplectic torus bundle with nontrivial monodromy and Euler numbers. The monodromy is given as a representation of the fundamental group of the symplectic base manifold of the supermembrane in the  homotopic  $ {\pi_0} $ group of symplectomorphisms of the fiber.
 In the case we will consider, the latest becomes the SL(2,Z) group acting on the fiber. It defines an automorphism on the fibers providing the global structure of the geometrical setting. 
The proof involves a nontrivial step in showing that the action of the Supermembrane with central charges, which explicitly depends on the moduli of the fiber manifold, is invariant under the Z-module associated to the monodromy. 

We think that this global construction, which allows a classification of all supermembranes that can be formulated as symplectic torus bundles with nontrivial monodromy and Euler numbers, is the origin in M-theory of the gauging of the effective theories associated to String theories \cite{sugra}.

The Supermembrane with nontrivial central charges, motivated by the light cone gauge formulation of the Supermembrane in \cite{hoppe},\cite{dwhn}, \cite{dwmn},\cite{dwln}, introduces a topological restriction on the physical configurations. It defines an associated Chern number.  From an algebraic point of view it can be interpreted as a nontrivial central charge in the Supersymmetric algebra. From a geometrical point of view it ensures the existence of a U(1) principle bundle and a monopole connection \cite{mr} on it whose curvature is in the Chern class associated to the topological restriction. In this sense, it is a natural way to introduce monopole configurations which stabilizes the supermembrane. In fact, the resulting regularized Hamiltonian has a discrete spectrum , that is the essential spectrum is  empty\cite{bgmr},\cite{bgmr2},\cite{br},\cite{bgmr3}. 
The additional global structure we will consider involves in a manifest way the SL(2,Z) duality group of String theory. Its supermembrane origin was emphasized in \cite{schwarz} in relation with the (p,q) string solutions, see also \cite{sl2z}. The SL(2,Z) group acts on the fiber bundle structure  as the cero homotopic group of symplectomorphisms preserving the symplectic form. This action induces a modular transformation on the basis of homology of the fiber, and correspondingly a Mobius transformation on the moduli of Teichmuller space. 
The final consistency in the construction arises when the global SL(2,Z) structure becomes compatible with the monopole ( or central charge) topological restriction.
\\

The paper is structured as follows. In section 2 we show some
properties of  a symplectic torus
bundle. In section 3 we present the local structure of the supermembrane  with
central charges. In section 4 we show how the hamiltonian of the
supermembrane with central charges is invariant under the group of
monodromies of the symplectic torus bundle and consequently, it may be formulated in terms of sections of symplectic torus bundles. Section 5 is devoted to conclusions.

\section{Supermembranes  and symplectic torus bundles}

A symplectic torus-bundle \cite{khan} is a smooth fiber bundle

 \bea\label{11}
 \xi:F\to E\stackrel{p}{\rightarrow} \Sigma
 \eea
 $F$ is the fiber, which we take to be the
2-torus $T^2$ . $E$ is the total space  and $\Sigma$  the base manifold which we
consider to be a closed, compact Riemann surface.  The structure group is the group of
symplectomorphisms preserving a given symplectic structure on $T^2$. So far, this
symplectic fibration naturally fits in a Supermembrane formulation in the light cone
gauge since this one is invariant under area preserving diffeomorphisms  which are
symplectomorphisms preserving the associated symplectic structure. The latest are
symplectomorphisms on the base manifold , however in the supermembrane with nontrivial
central charges \cite{torrealba} which we will describe in detail in the following sections , these
ones correspond to the pull-back of the symplectomorphisms on the fiber. In order to
describe global aspects of the supermembrane we introduce a monodromy and the associated 
Z-module. The monodromy is the natural extension of the monodromy in a  torus bundle on a
circle as considered  by Thurston. We follow here the approach of \cite{khan}. Related work may be found in \cite{geiges}, \cite{sakamoto-fukuhara}, \cite{walzack} and \cite{thurston}.

The action of the structure group on $T^2$ produces a $\pi_0(G)$
action on the homology and cohomology groups of $T^2$.
The homomorphisms $\pi_1(\Sigma)\to \pi_0(G)$ give to each homology and cohomology
group of $T^2$, the structure of $Z[\pi_1(\Sigma)]$-module.
$\pi_0(G)$ in the case under consideration  is known to be $SL(2,Z)$. Moreover, the action of
$\pi_0(G)$ on $H_1(T^2)$, the first homology group, may be identified
with the natural action of $SL(2,Z)$ on $Z ^2$. Given any
representation $\rho:\pi_1(\Sigma)\to SL(2,Z)$ we denote $Z_{\rho}^2$ the
corresponding $Z[\pi_1(\Sigma)]$-module.

A theorem, see\cite{khan},ensures the existence of a bijective
correspondence between the equivalent classes of the symplectic
torus bundle together with a representation $\rho$ inducing the
module structure $Z_{\rho}^2$ on $H_1(T^2)$ and the elements of
$H^2(\Sigma,Z_{\rho}^2)$, the second cohomology group of $\Sigma$ with
local coefficients $Z_{\rho}^2$. This theorem classifies the
symplectic torus bundle $\xi$ with a representation $\rho$ in terms
of the characteristic class $C(\xi)$. In order to formulate
the supermembrane with central charges in terms of sections   of a
symplectic torus bundle with a representation $\rho$  inducing a
$Z[\pi_1(\Sigma)]$-module, we have to consider the transformation of
its hamiltonian under the action of $SL(2,Z)$ on the homology basis
since the moduli of the 2-torus $T^2$ appear explicitly in the
hamiltonian.

\section{The supermembrane in the Light Cone Gauge}

In this section we describe the Supermembrane with non trivial central charges in terms of maps from the base manifold to the target space. It corresponds to a formulation in terms of local sections of a symplectic torus bundle.

The hamiltonian of the $D=11$ Supermembrane [1] may be defined in
terms of maps $X^{\mu}$, $\mu = 0, . . . , 10$, from a base manifold
$\Sigma\times R$ onto a target manifold which we will assume to be
11D Minkowski. $\Sigma$ is a Riemann surface of genus g.
$\sigma^a$,$a=1,2$ are local spatial coordinates over $\Sigma$ and
$\tau\in R$ represents the worldvolume time. Decomposing $X^{\mu}$
and $P_{\mu}$ accordingly to the standard Light Cone Gauge
(LCG) ansatz and solving the constraints, the canonical reduced
hamiltonian of the $D=11$ supermembrane is given by \bea
H=T^{-2/3}\int_{\Sigma}\sqrt{W}\left[\frac{1}{2}\left(\frac{P_M}{\sqrt{W}}\right)^2+\frac{T^2}{4}\{X^{M},X^{N}\}^2
+\sqrt{W}\overline{\theta}\Gamma_{-}\Gamma_m\{X^m,\theta\}\right]
\eea subject to the constraint \bea\label{1} \phi\equiv
d(P_MdX^M+\overline{\theta}\Gamma_{-}\theta)=0 \eea and to the
global one \bea\label{2}
\phi_0\equiv\int_{\mathcal{C_s}}P_MdX^M+\overline{\theta}\Gamma_{-}d\theta=0
\eea where $\mathcal{C}_s$ is a basis of homology on $\Sigma$, with
$M=1,\dots,9$, and $P_M$ are the conjugate momenta to $X^M$.
$\sqrt{W}$ is the scalar density introduced in the LCG ,
$\Sigma$ is the base manifold which we take to be a
Riemann surface, $\theta$ represents the 11D Majorana spinors and $\Gamma_{\mu}$ are the corresponding Dirac matrices. $T$ is the tension of the supermembrane. $\phi$ and $\phi_0$ are the
generators of the area preserving diffeomorphims homotopic to the
identity and they preserve the area element
$\sqrt{W}\epsilon_{ab}d\sigma^a\wedge d\sigma^b$, a symplectic 2-form.

\bea\{X^m,X^n\}=\frac{\epsilon^{ab}}{\sqrt{W}}\partial_aX^m\partial_bX^n\eea
is the associated symplectic bracket. We now consider the
supermembrane wrapped on a compact sector of the target space,
restricted by a topological condition: the supermembrane with
nontrivial central charges. 

\subsection{The supermembrane with nontrivial central charges}

We consider the Supermembrane on a target space $M_9\times T^2$  where
$T^2$ is a flat torus defined in terms of a lattice $\mathcal{L}$ on
the complex plane $C$:
 \bea\label{3}
 \mathcal{L}: z\to z+2\pi R(l+m\tau),
 \eea
where $m,l$ are integers, R is a real moduli, $R>0$, and $\tau$ a
complex moduli $\tau=Re\tau +iIm\tau$, $Im\tau>0$, $T^2$ is defined
by ${C} / {\mathcal{L}}$. $\tau$ is the complex coordinate of the
Teichmuller space for $g=1$, that is the upper half plane. The
Teichmuller space is a covering of the moduli space of Riemann
surfaces, it is a $2g-1$ complex analytic simply connected manifold
for genus $g$ Riemann surfaces.

The conformally equivalent tori are identified by the parameter
$\tau$ modulo the Teichmuller modular group, which in the case $g=1$
is $SL(2,Z)$.   It acts on
the Teichmuller space through a Mobius transformation and it has a
natural action on the homology group $H_1(T^2)$. In order to
define a supermembrane with nontrivial central charges we consider
maps $X^m,X^r$  from $\Sigma$ to the target space , with $r=1,2;
 m=3,\dots,9$ where $X^m$ are single valued maps onto the Minkowski sector of the target
space while $X^r$ map onto the $T^2$ compact sector of the
target. The necessary winding conditions, in order to define a
map onto the $T^2$, are:
 \bea\label{4}
 \oint_{\mathcal{C}_s}dX=2\pi R(l_s+m_s\tau)
 \eea
where $dX=dX^1+idX^2$, $l_s$ and $m_s$ , $s=1,2$ are integers and
$\mathcal{C}_s$ the above mentioned basis of homology for a genus $g=1$ Riemann surface. From now on, $\Sigma $ will be a Riemann surface of  genus $g=1$. We will denote $d\widehat{X}^r, r=1,2$ a normalized
basis of harmonic one forms on $\Sigma$:
 \bea\label{55}
 \oint_{\mathcal{C}_s}d\widehat{X}^r=\delta_r^s
 \eea

We may decompose the closed one-forms $dX^r$ in terms of harmonic
one forms plus exact ones. We obtain from (\ref{4})
 \bea\label{6}
 dX=2\pi R (l_s+m_s\tau)d\widehat{X}^s+dA
 \eea
where $dA$ denotes the exact one-form. We now impose a topological
restriction on the winding maps: the irreducible winding constraint,
 \bea\label{7}
 \int_{\Sigma}dX^r\wedge dX^s=n\epsilon^{rs}Area(T^2)\quad r,s=1,2
 \eea
where the winding number $n$ is assumed to be different from zero.
$\epsilon^{rs}$ is the symplectic antisymmetric tensor associated to
the symplectic 2-form on the flat torus $T^2$. In the case under
consideration $\epsilon^{rs}$ is the Levi Civita antisymmetric
symbol. An important point implied by the assumption $n\ne 0$ is
that the cohomology class in $H^2(\Sigma,Z)$ is non-trivial.

It also implies that there is an $U(1)$ nontrivial principle bundle
over $\Sigma$ and a connection on it whose curvature is given by
$d\widehat{X}^r\wedge d\widehat{X}^s$. This $U(1)$ nontrivial
principal fiber bundle are associated to the presence of monopoles
on the worldvolume of the supermembrane explicitly discussed in \cite{mr}.

The natural scalar density $\sqrt{W}$ on the geometrical picture we
are considering is obtained from the pullback of the symplectic 2-form on $T^2$ by the map $\widehat{X}^r, r=1,2,$
   \bea\label{8}
   W=\epsilon_{rs}d\widehat{X}^r\wedge d\widehat{X}^s\equiv \sqrt{W}\epsilon_{ab}d\sigma^a\wedge d\sigma^b
   \eea
where
$\sqrt{W}=\frac{1}{2}\epsilon_{rs}\partial_a\widehat{X}^r\partial_b\widehat{X}^s\epsilon^{ab}$.

The symplectomorphisms preserving the canonical symplectic structure  on $T^2$ are then  pull-back to symplectomorphisms preserving $W$ on $\Sigma$.
This is relevant in the construction of the supermembrane with central charges as sections of a symplectic torus bundle.
 Using $Area(T^2)=(2\pi R)^2 Im\tau$, condition (\ref{7}) implies
 \bea\label{9}
 n=det \begin{pmatrix} l_{1}& l_{2}\\
                       m_{1} & m_{2}
 \end{pmatrix}
 \eea
That is, all integers $l_s,m_s, s=1,2$ are admissible provided they
satisfy restriction (\ref{9}).

The supermembrane with non-trivial central charges is invariant
under area preserving diffeomorphims homotopic to the identity. In
particular, under conformal maps which leave invariant the homology
basis on $\Sigma$. In fact, $d\widehat{X}^r$ remain invariant and
hence the symplectic 2-form in $\Sigma$. It is also invariant
under  diffeomorphisms not homotopic to the identity acting on the homology
basis in $\Sigma$ as $SL(2,Z)$ transformations. In fact if
$d\widehat{X}^r(\sigma)\to S_r^s d\widehat{X}^r(\sigma)$  with
$[S_s^r]\in SL(2,Z)$, then the symplectic 2-form $W$ remains
invariant. In this case
 \bea\label{100}
 dX\to 2\pi R (l_s +m_s\tau)S_r^s d\widehat{X}^r+dA
 \eea
where the exact part $A$ transform as a scalar field. Consequently, if we also transform the winding integers by
\begin{equation}\label{10}
     \begin{pmatrix} l_{1}& l_{2}\\
                       m_{1} & m_{2}
 \end{pmatrix}\to\begin{pmatrix} l_{1}& l_{2}\\
                       m_{1} & m_{2}
 \end{pmatrix}\begin{pmatrix} S_{1}^1& S_{2}^1\\
                       S_{1}^2 & S_{2}^2
 \end{pmatrix}
\end{equation}
then the harmonic part of $dX$ remains invariant. The $SL(2,Z)$
acts from the right on the winding matrix.

\section{The supermembrane maps as sections of a symplectic torus bundle}

In this section we are going to prove the invariance of the
Hamiltonian under the $Z[\pi_1(\Sigma)]$-module. The hamiltonian of
the supermembrane with central charges is given by \bea\label{5}
H=\int_{\Sigma}\mathcal{H}=&
\int_{\Sigma}T^{-2/3}\sqrt{W}[\frac{1}{2}(\frac{P_{m}}{\sqrt{W}})^{2}+
\frac{1}{2}(\frac{P_{r}}{\sqrt{W}})^{2}
+\frac{T^{2}}{2}\{X^{r},X^{m}\}^{2}\\ \nonumber &
+\frac{T^{2}}{4}\{X^{r},X^{s}\}^{2}+\frac{T^{2}}{4}\{X^{m},X^{n}\}^{2}]+\textrm{fermionic
terms} \eea subject to (\ref{1}),(\ref{2}), and (\ref{4}),
(\ref{7}), where $X^r$ are sections on the symplectic torus bundle
$\xi$ with structure group $G$, the symplectomorphims preserving the
symplectic 2-form on the fibre  $T^2$ defined previously . $P_r$ are the
conjugate momenta to the exact part in the decomposition of $X^r$.
The integrand depending in $X^r, r=1,2$ may be re-written in terms
of \bea\label{12} dX=dX^1+idX^2 \eea as \bea\label{13}
\frac{1}{2}\{X,X^m\}\{\overline{X},X^m\}+\frac{1}{8}\{X,\overline{X}\}\{\overline{X},X\},
\eea where \bea\label{14} dX=2\pi R (l_s+m_s\tau)d\widehat{X}^s+dA
\eea where $R$ and $\tau$ are the moduli of the $T^2$,
$d\widehat{X}^s,s=1,2$ as before are the harmonic basis of $\Sigma$
and $dA$ is the exact one-form in the Hodge decomposition.
$A=A_1+iA_2$ carries the physical degrees of freedom of the compact
sector. The action of $\pi_0(G)\equiv SL(2,Z)$ in $H_1(T^2)$ is the
natural one inducing a Mobius transformation on the upper half plane
with complex coordinate $\tau$.  We now prove
that the hamiltonian (\ref{5}) is a well defined functional on the
symplectic torus bundle with monodromy $\rho$, where $\rho$ is a representation of  $\pi_1(\Sigma)$ in $SL(2,Z)$ as defined in section 2. In fact, it is
invariant under the following transformation on $T^2$:
\bea \label{17}\tau &\to& \frac{a\tau+b}{c\tau+d}\\
\nonumber  R& \to & R|c \tau +d| \\ \nonumber  A &\to& A
e^{i\varphi_{\tau}}
\\
\nonumber  \begin{pmatrix} l_{1}& l_{2}\\
               m_{1} & m_{2}
 \end{pmatrix}&\to& \begin{pmatrix} a & -b\\
                        -c & d
 \end{pmatrix}\begin{pmatrix} l_{1}& l_{2}\\
               m_{1} & m_{2}
 \end{pmatrix}\eea

where $c\tau+d= |c\tau+d|e^{-i\varphi_{\tau}}$ and $\begin{pmatrix} a & b\\
                        c & d
 \end{pmatrix}\in Sp(2,Z)$.

this invariance was found in \cite{sl2z}. The hamiltonian (\ref{5}) as well as (\ref{7}), (\ref{4}) and the $Area(T^2)$ are invariant under the above transformation.

Notice that the $SL(2,Z)$ (\ref{17}) acts from the left on $\begin{pmatrix} l_{1}& l_{2}\\
               m_{1} & m_{2}
 \end{pmatrix}$
while the $SL(2,Z)$ invariance on the basis $\Sigma$, discussed in previous sections, acts on the right.

Under these transformations the $det\begin{pmatrix} l_{1}& l_{2}\\
               m_{1} & m_{2}
 \end{pmatrix}$ remains invariant. Given $\begin{pmatrix} l_{1}& l_{2}\\
               m_{1} & m_{2}
 \end{pmatrix}$
 with determinant $ \ne 0$ there always exist elements of $SL(2,Z)$ whose action from the left and from the right yields
 \bea\label{18}
\begin{pmatrix} a & b\\
                        c & d
 \end{pmatrix} \begin{pmatrix} l_{1}& l_{2}\\
               m_{1} & m_{2}
 \end{pmatrix} \begin{pmatrix} S^{1}_{1}& S^{1}_{2}\\
                       S^{2}_{1} & S^{2}_{2}

 \end{pmatrix}=\begin{pmatrix} \lambda_1& 0 \\
                       0 & \lambda_2
 \end{pmatrix}
 \eea
where $\lambda_1\lambda_2=n$.  Moreover, if $\lambda_1$ and
$\lambda_2$ are relative primes there always exist elements
belonging to $SL(2,Z)$  whose action from the left and the right yield
$\lambda_1=n$ and $\lambda_2=1$. If $\lambda_1$ and $\lambda_2$ are
not relative primes one may redefine the parameter $R$ and reduce
to the case where $\lambda_1$ and $\lambda_2$ are relative primes.
 We thus obtain a canonical expression for the hamiltonian (\ref{5}) subject to (\ref{1}),(\ref{2}), and (\ref{4}),
(\ref{7}), in terms of sections of the symplectic torus bundle with a
monodromy $\rho$ : \bea\label{5a}
H=&\int_{\Sigma}\mathcal{H}=\int_{\Sigma}T^{-2/3}\sqrt{W}\left[\frac{1}{2}(\frac{P_{m}}{\sqrt{W}})^{2}+
\frac{1}{2}(\frac{P_{r}}{\sqrt{W}})^{2}
+\frac{T^{2}}{2}\{{X},X^m\}\{\overline{X},X^m\}\right]\\ \nonumber &
+\left[\frac{T^{2}}{8}\{X,\overline{X}\}\{\overline{X},X\}+\frac{T^{2}}{4}\{X^{m},X^{n}\}^{2}\right]+\textrm{fermionic
terms} \eea
where $dX= 2\pi R (d\widehat{X}^1+n\tau d\widehat{X}^2)$. Although
we may have winding numbers $l_1,l_2,m_1,m_2$ the symmetries of the
theory allow to reduce everything to the central charge integer $n$.

The final point is to determine which representations
$\rho:\pi_1(\Sigma)\to\pi_0(G)\equiv SL(2,Z)$ leave invariant the
form of the hamiltonian density in (\ref{5a}). The representations
$\rho_n:\pi_1(\Sigma)\to SL(2,Z)_n$, where $SL(2,Z)_n$ is the
subgroup of $SL(2,Z)$ whose elements are of the form
 \bea\label{180}
\begin{pmatrix} a & nb\\
                        c & d
 \end{pmatrix}
 \eea
 leave invariant the hamiltonian density in (\ref{5a}). $\rho_n$ characterizes the representations compatible with the topological restriction (\ref{7}) For example, if we take the representation $\rho:\pi_1(\Sigma)\to SL(2,Z)_n$ defined in the following way:
 \bea\label{181}
 \pi_1(\Sigma)\ni \begin{pmatrix} M\\
                        N
 \end{pmatrix}\rightarrow \begin{pmatrix} 1 & nM\\
                        0& 1
 \end{pmatrix}
 \eea
 The element of $H_1(T^2)$ may be given by $\begin{pmatrix} p\\
                    q
 \end{pmatrix}$  being $p,q$ integers. Then the natural action of $SL(2,Z)$ on it is given by
  \bea\label{182}
\begin{pmatrix} 1 & nM\\
                        0& 1
 \end{pmatrix}\begin{pmatrix} p\\
                       q
 \end{pmatrix}=\begin{pmatrix} p+nMq\\
                       q
 \end{pmatrix}
 \eea
 The cohomology group $H^2(\Sigma,Z^2_{\rho})\cong\mathbb{Z}$, also the central charge condition (\ref{7}) states that we are in the characteristic class $C(\xi)=n\ne 0$, consequently, there exists a $D=11$ supermembrane with nontrivial central charges formulated in terms of sections of a symplectic torus bundle $\xi$ with representation (\ref{181}) inducing a $Z[\pi_1(\Sigma)]$-module. 
\section{Conclusion}

We showed that the Supermembrane with central charges may be
formulated in terms of sections of symplectic torus bundles with a
representation $\rho:\pi_1(\Sigma)\to SL(2,Z)$ inducing a
$Z[\pi_1(\Sigma)]$-module in terms of the $H_1(T^2)$ homology group
of the fiber. The representation $\rho$ may be interpreted as a
monodromy on the bundle. The non trivial point in the construction
was to prove that the hamiltonian together with the constrains are
invariant under the action of $ SL(2,Z)$ on the homology group $H_1(T^2)$ of
the fibre 2-torus $T^2$. An interesting aspect of this geometrical structure is the possible
existence of an extension of the symplectic 2-form on the
fiber to the full space of the symplectic torus bundle. A theorem of
Khan \cite{khan} establishes that the extension exists if and only
if the characteristic class is a torsion class in
$H^2(\Sigma,Z_{\rho}^2)$. In the case of the example of section 4, we
conclude that there is not such extension since $C(\xi)=n$ is not a
torsion class. The only one is $C(\xi)=0$ which is not compatible
with the topological restriction (\ref{7}) of the supermembrane with
central charges.  Locally we have the usual
interpretation of the supermembrane in terms of maps from $\Sigma$
to the target. Globally we have now a more interesting geometrical
structure since the hamiltonian is defined on a non-trivial
symplectic torus bundle. Locally the target is a product of
$M_9\times T^2$ but globally we cannot split the target from the
base $\Sigma$ since $T^2$ is the fiber of the non trivial symplectic torus
bundle $T^2\to\Sigma$. The formulation of the supermembrane in terms of
sections of the symplectic torus bundle with a monodromy  is a nice geometrical structure to analyze global aspects of gauging procedures on effective theories arising from M-theory. We noticed the particular case in which the representation $\rho$ is given by the matrix 
 \bea\label{182}
\begin{pmatrix} 0 & 1\\
                        -1& 0
 \end{pmatrix}^{M+N}
 \eea 
 the subgroup reduces to $Z_2\times Z_2$ and this case was considered in \cite{ovalle}-\cite{g2}.

\section{Acknowledgements}

We would like to thank for interesting comments to prof. A.
Lozada and  we are specially indebted for enlighting clarifications to Prof.
A. Vi\~na.  The work of MPGM is funded by the Spanish Ministerio de
Ciencia e Innovaci\'on (FPA2006-09199) and the Consolider-Ingenio
2010 Programme CPAN (CSD2007-00042). IM and AR acknowledge  support from DID-USB. AR acknowledge to the Physics Department of UA (Chile) 
for financial support.


\begin{thebibliography}{99}

\bibitem{ho} Pei-Ming Ho {\em Twisted bundle on quantum torus and BPS states in matrix theory.}
Phys.Lett.{\bf B434}:41-47,1998.
e-Print: hep-th/9803166

\bibitem{ovrut} B. A. Ovrut, T. Pantev, R. Reinbacher, {\em Torus fibered Calabi-Yau threefolds with nontrivial fundamental group.} JHEP {\bf 0305}:040,2003.
e-Print: hep-th/0212221

\bibitem{ralph} R. Blumenhagen, B. Jurke, T. Rahn, H. Roschy, {\em Cohomology of Line Bundles: Applications.}
e-Print: arXiv:1010.3717 [hep-th] 

\bibitem{3hull99} C. M. Hull, {\em Duality and strings, space and
time}. Contribution to Proceedings of "Mathematical Sciences beyond
the Second Millenium" held at the Center for Advanced Mathematical
Sciences, Beirut. {\tt hep-th/9911080}.

\bibitem{4hull04} C. M. Hull, {\em A geometry for non-geometric string
backgrounds}. JHEP 0510:065, 2005. {\tt hep-th/0406102}.

\bibitem{5wetch}
J. Shelton, W. Taylor, B. Wecht, {\em Nongeometric flux
compactifications}. JHEP 0510:085, 2005. {\tt hep-th/0508133}.

\bibitem{6hullreid} C.M. Hull, R. A. Reid-Edwards, {\em Flux compactifications of
string theory on twisted tori}. {\tt hep-th/0503114}.

\bibitem{7samtleben} H. Samtleben, {\em Lectures on Gauged Supergravity and Flux
Compactifications}. Class. Quant. Grav. 25 (2008) 214002. {\tt
hep-th/0808.4076}.

\bibitem{11dabholkarhull} A. Dabholkar, C. Hull, {\em Duality twists, orbifolds, and
fluxes}. JHEP 0309:054, 2003. {\tt hep-th/0210209}.

\bibitem{12flournoyw} A. Flournoy, B. Williams, {\em Nongeometry, duality twists, and
the worldsheet}. JHEP 0601:166, 2006. {\tt hep-th/0511126}.

\bibitem{13kawai} S. Kawai and Y. Sugawara, {\em D-branes in T-fold conformal field
theory}. JHEP 0802:027, 2008. {\tt hep-th/0709.0257}.

\bibitem{reids}R A Reid-Edwards, {\em Flux compactifications, twisted tori and doubled
geometry}. JHEP 0906:085, 2009. {\tt hep-th/0904.0380}.

\bibitem{17hu8ll-reids}C. M. Hull and R. A. Reid-Edwards, {\em Flux compactifications of
M-theory on twisted tori}. JHEP 0610:086, 2006. {\tt
hep-th/0603094}.

\bibitem{18cederwall} M. Cederwall, {\em M-branes on U-folds}. {\tt
hep-th/0712.4287}.

\bibitem{20hull-mtheory} C. M. Hull, {\em Generalised geometry for M-theory}. JHEP 0707:079, 2007. {\tt
hep-th/0701203}.

\bibitem{24hull-tfolds} C. M. Hull, {\em Global Aspects of T-Duality, Gauged Sigma Models and
T-Folds}. JHEP 0710:057, 2007. {\tt hep-th/0604178}.

\bibitem{hoppe}] J. Hoppe {\em Two problems in quantum mechanics} Massachusetts Institute Of
Technology.Ph.Thesis. 1980. M.S.


\bibitem{dwhn} B. de Wit, J. Hoppe, H. Nicolai, {\em On the quantum mechanics of
supermembranes}. Nucl. Phys. {\bf B305}: 545,1988.

\bibitem{dwln} B. de Wit, M. Luscher, H. Nicolai, {\em The supermembrane is unstable}.
Nucl. Phys. {\bf B320}: 135, 1989.

\bibitem{dwmn} B. de Wit, U. Marquard, H. Nicolai,
{\em Area preserving diffeomorphisms and supermembrane lorentz
invariance}. Commun. Math. Phys. {\bf 128}: 39-62, 1990.

\bibitem{hulldabholkar} A. Dabholkar, C. Hull, {\em Generalised T-duality and
non-geometric backgrounds}. JHEP 0605:009, 2006. {\tt
hep-th/0512005}.

\bibitem{hullreid} C.M. Hull, R. A. Reid-Edwards, {\em Flux
compactifications of string theory on twisted tori}. Fortsch. Phys.
{\bf 57}: 862-894, 2009. {\tt hep- th/0503114}.

\bibitem{sugra} M.P. Garcia del Moral, J.M. Pena, A. Restuccia. In preparation.

\bibitem{gmr} M.P. Garcia del Moral, A. Restuccia, {\em On the
spectrum of a noncommutative formulation of the D=11 supermembrane
with winding}. Phys. Rev. {\bf D66} 045023, 2002. {\tt
hep-th/0103261}.

\bibitem{dwpp} B. de Wit, K. Peeters, J. Plefka,
{\em Supermembranes with winding}. Phys. Lett. {\bf B409}: 117-123,
1997. {\tt hep-th/9705225}

\bibitem{bgmmr} L. Boulton, M. P. Garcia del Moral, I.
Martin, A. Restuccia, {\em On the spectrum of a matrix model for the
D=11 supermembrane compactified on a torus with non-trivial
winding}. Class. Quant. Grav. {\bf 19}: 2951, 2002. {\tt
hep-th/0109153}

\bibitem{torrealba} I. Martin, A. Restuccia, R. S. Torrealba,
{\em On the stability of compactified D = 11 supermembranes}.
 Nucl. Phys. {\bf B521}: 117-128, 1998. {\tt hep-th/9706090}.

\bibitem{bgmr} L. Boulton, M.P.
Garcia del Moral, A. Restuccia, {\em Discreteness of the spectrum of
the compactified D=11 supermembrane with non-trivial winding}. Nucl.
Phys. {\bf B671}: 343-358, 2003. {\tt hep-th/0211047}.

\bibitem{schwarz} J. H. Schwarz, {\em An SL(2,Z) multiplet of type IIB superstrings.}
 Phys.Lett.{\bf B360}:13-18,1995, Erratum-ibid.B364:252,1995.
e-Print: hep-th/9508143


\bibitem{bgmr2}L. Boulton, M.P. Garcia del Moral, A. Restuccia
{\em The Supermembrane with central charges: (2+1)-D NCSYM,
confinement and phase transition}. Nucl. Phys. {\bf B795}: 27-51,
2008. {\tt hep-th/0609054}.

\bibitem{br} L. Boulton, A. Restuccia {\em The heat kernel of the compactified D=11 supermembrane with non-trivial winding}
     Nucl.Phys. {\bf B724} (2005) 380-396 {\tt arXiv:hep-th/0405216}
     
\bibitem{mr} I. Martin, A. Restuccia, {\em Magnetic monopoles over topologically non trivial Riemann
Surfaces}. Lett. Math. Phys. {\bf 39} 379-391, 1997.  {\tt
hep-th/9603035}.
 
 \bibitem{sl2z} M.P. Garcia del Moral, I. Martin, A. Restuccia, {\em Nonperturbative SL(2,Z) (p,q)-strings manifestly realized on the quantum
M2}. {\tt hep-th/0802.0573}.
    
\bibitem{ovalle}I. Martin, J. Ovalle, A. Restuccia, {\em D-branes, symplectomorphisms and noncommutative gauge theories.}
Nucl. Phys. Proc. Suppl. {\bf 102}: 169-175, 2001; {\em Compactified
D = 11 supermembranes and symplectic noncommutative gauge theories}.
Phys. Rev. {\bf D64}: 046001, 2001. {\tt hep-th/0101236}.


\bibitem{bgmr3} L. Boulton, M. P. Garcia del Moral, A. Restuccia, {\em Spectral properties in supersymmetric matrix models}.
{\tt hep-th/1011.4791}.

\bibitem{bellorin} J. Bellorin, A. Restuccia,
{\em D=11 Supermembrane wrapped on calibrated submanifolds}. Nucl.
Phys. {\bf B737}: 190-208, 2006. {\tt hep-th/0510259}.


\bibitem{joselen} M. P. Garcia del Moral, J. M. Pena, A. Restuccia, {\em N=1 4D Supermembrane from
11D}. JHEP 0807:039, 2008. {\tt hep-th/0709.4632}.

\bibitem{g2} A. Belhaj, M.P. Garcia del Moral, A. Restuccia, A. Segui, J.P. Veiro {\em The Supermembrane with Central Charges on a G2 Manifold}
J.Phys.{\bf A42}:325201,2009 arXiv:0803.1827  
    
\bibitem{geiges} H. Geiges, {\em Symplectic structures on T2-bundles over T2}. Duke Math. J. Volume {\bf 67}, Number 3 (1992), 539-555.

\bibitem{sakamoto-fukuhara} K. Sakamoto, S. Fukuhara, {\em Classification of
$T^2$-bundles over $T^2$}. Tokyo J. Math. {\bf 6}: 311-327, 1983.

\bibitem{walzack} R. Walczak, {\em Existence of symplectic structures on torus bundles over surfaces}. {\tt
math/0310261}

\bibitem{khan} P. J. Khan, {\em symplectic torus bundles and group extensions}. New York Journal of Mathematics. New York J. Math. {\bf 11}: 35–55, 2005.

\bibitem{thurston} W. Thurston, {\em Some simple examples of symplectic manifolds}. Proc. of the AMS 55 (1976),
467–8.

\bibitem{kravchenko} O. Kravchenko, {\em Deformation Quantization of Symplectic Fibrations}. Compositio Mathematica
Volume 123, Number 2, 131-165, DOI: 10.1023/A:1002452002677.


\end{thebibliography}
\end{document}